\documentclass[aps,reprint,superscriptaddress,nofootinbib,preprintnumbers,longbibliography]{revtex4-2}

\usepackage{standalone}

\usepackage{url}
\usepackage{upgreek,amssymb,multirow}
\usepackage{mathtools}
\usepackage{float}
\usepackage{mciteplus}
\usepackage{blindtext}

 \usepackage{dsfont}

\usepackage{comment}

\usepackage{amsmath}
\usepackage{extarrows}
\usepackage{braket}

\usepackage[dvipsnames,svgnames]{xcolor}

\usepackage{tikz-cd}
\usetikzlibrary{
  calc,
  topaths,
  decorations,
  decorations.pathmorphing,
  arrows,
  decorations.markings,
  cd,
  positioning,
  shapes.geometric
}
\tikzset{
->-/.style args={#1rotate#2}{decoration={markings, mark=at position #1 with {\arrow[scale=1.5,rotate = #2 ]{stealth}}}, postaction={decorate}}
}

\tikzset{line/.style={line width=0.25mm},
curve/.style={line,smooth,tension=1},
->-/.style={decoration={
  markings,
  mark=at position #1 with {\arrow[>=stealth]{>}}},postaction={decorate}},
-<-/.style={decoration={
  markings,
  mark=at position #1 with {\arrow[>=stealth]{<}}},postaction={decorate}},
}

\usepackage{breakurl}
\usepackage[hyperindex,breaklinks]{hyperref}

\newcommand{\ie}{\begin{equation}\begin{aligned}}
\newcommand{\fe}{\end{aligned}\end{equation}}

\newcommand{\D}{\mathsf{D}}

\newcommand{\cD}{{\mathcal{D}}}

\newcommand{\cH}{{\cal H}}

\newcommand{\cC}{{\cal C}}

\newcommand{\ra}{\rightarrow}

\newcommand{\Vc}{{\rm Vec}}
\newcommand{\Rep}{{\rm Rep}}

\newcommand{\End}{{\rm End}}

\newcommand{\CC}{{\mathbb C}}
\newcommand{\ZZ}{{\mathbb Z}}

\newcommand{\cO}{\mathcal O}

\begin{document}

\title{Symmetry-permuting entanglers for non-invertible symmetry-protected topological phases}
 
\author{Minyoung You}
\email{miyou849@gmail.com}
\affiliation{ Yukawa Institute for Theoretical Physics, Kyoto University, Kitashirakawa Oiwakecho,
 Sakyo-ku, Kyoto 606-8502, Japan}

\begin{abstract}
1+1D symmetry-protected topological (SPT) phases with invertible
symmetry can be characterized  by the symmetric entanglers: globally symmetric finite-depth local unitary circuits that generate representative ground states from a product state.  For non-invertible
symmetries,  however, distinct SPT phases, such as the three $\mathrm{Rep}(D_8)$ SPT phases, may not be connected
by any symmetric entangler. We show that the three
$\mathrm{Rep}(D_8)$ SPT phases can nevertheless be connected by
\emph{symmetry-permuting} finite-depth local unitary circuits,  realizing  dualities predicted by topological quantum field theory (TQFT). We construct  an explicit
 matrix product unitary connecting two of the $\mathrm{Rep}(D_8)$ SPT phases, and show that its action on string order parameters reproduces the anyon permutation predicted by the TQFT duality. 
\end{abstract}


\maketitle

\tableofcontents

\section{Introduction}

A useful way to characterize a one-dimensional symmetry-protected topological (SPT)
phase is through a finite-depth local unitary circuit that generates it from a product
state. For ordinary invertible symmetries, such an entangler can be chosen to preserve
the symmetry globally even though its individual local gates need not be symmetric
\cite{chen2012symmetry,Chen_2013,zeng2018quantuminformationmeetsquantum,tantivasadakarn2023pivot}. In this setting, classifying SPT phases and
classifying the corresponding symmetry-preserving entanglers lead to the same group
cohomology data \cite{zhang2023topological}.

For non-invertible symmetries, the correspondence is no longer so simple. A striking
example is provided by the three $1+1$D SPT phases with $\Rep(D_8)$ symmetry constructed
in Ref.~\cite{seifnashri2024cluster}: distinct phases cannot be connected by a
finite-depth circuit that commutes with the full non-invertible symmetry. By contrast,
a symmetric entangler was constructed for the two $\Rep(A_4)$ SPT phases in
Ref.~\cite{You_2026}. These examples indicate that non-invertible SPT phases can differ not
only by their phase labels, but also by the type of finite-depth duality, if any, that
connects them.

This distinction is naturally anticipated by the symmetry topological quantum field theory (symTFT) description. In that
framework, dualities of the boundary theory are encoded by braided autoequivalences of
the bulk Drinfeld center. Ref.~\cite{aksoy2025} distinguished, in particular, dualities
that preserve the Lagrangian algebra of charges, while either fixing or permuting its
simple summands. The former give fixed-charge dualities (FCDs), while the latter give
fixed-algebra dualities (FADs). This suggests a microscopic resolution of the apparent
$\Rep(D_8)$ obstruction: the absence of a symmetry-\emph{commuting} entangler need not
imply the absence of a finite-depth entangler. Instead, the appropriate circuit may  permute its symmetry lines.

In this work we develop this idea concretely. First, in Sec. \ref{sec:symmetry-permuting-entanglers}, we construct
explicit symmetry-permuting entanglers for the $\Rep(D_8)$ SPT phases. The non-on-site
generator is represented by a bond-dimension-two matrix product unitary (MPU). Unlike
the earlier $\Rep(A_4)$ construction, whose local tensor was found essentially ad hoc,
here the MPU is derived from local symmetry pulling-through constraints. Starting from
the action tensors of the product-state and SPT matrix product states, we determine the
crossing tensors, solve the resulting linear constraints on the MPU tensor, and then
impose exact state mapping, hermiticity, and unitarity conditions. The
resulting circuit exchanges two of the $\Rep(D_8)$ SPT phases and has strict order two.

As we show in Sec.~\ref{sec:anyon-permutations}, the constructed circuit realizes the corresponding FAD microscopically, in the sense that it permutes string order parameters in precisely the manner predicted by the FAD action on anyons. Combined with an on-site unitary realizing a second FAD arising from a group automorphism of $D_8$, these transformations generate the full $S_3$ group of FADs and connect all three SPT phases.
Our results therefore substantially strengthen the proposed microscopic interpretation of the
symTFT classification.

\section{Background: non-invertible SPT phases and two notions of dualities}
\label{sec:background}

\subsection{Fusion-category symmetry and the symmetry TQFT}

\begin{figure}[t]
    \centering
\raisebox{-73pt}{\begin{tikzpicture}
\draw [<->] (1,4.2) -- (3,4.2);
\draw (2,4.2) node[above]{$I$};
\draw (1.35,0.8) node[below]{\small \color{red!75!DarkGreen} $Z(\cC)$};
\draw[color=DarkGreen] (0,0) -- (0,3) -- (1,4) -- (1,1) -- cycle;
\draw (0,0) node[below]{\color{DarkGreen} $\mathbb{B}_Q$};
\draw [color=red!75!DarkGreen, thick, decoration = {markings, mark=at position 0.6 with {\arrow[scale=1]{stealth}}}, postaction=decorate] (0.5,2) -- (1.5,2) node[below]{$\mu$} -- (2.5,2);
\draw [fill=DarkGreen] (0.5,2) circle (0.04) node [below] {\color{DarkGreen} $V_\mu$};
\draw [color=blue!70!green, thick, decoration = {markings, mark=at position 0.5 with {\arrow[scale=1]{stealth}}}, postaction=decorate] (2.5,2) -- (2.5,3) node[right]{$a$} -- (2.5,3.75);
\draw [fill=blue!70!green] (2.5,2) circle (0.04) node [below] {\color{blue!70!green} $W_a^\mu$};
\draw[color=blue!70!green, preaction={draw=white,line width=3pt}] (2,0) -- (2,3);
\draw[color=blue!70!green] (2,3) -- (3,4) -- (3,1) -- (2,0);
\draw (2,0) node[below]{\color{blue!70!green} $\D$};
\end{tikzpicture}}
    \quad $=$ ~
\raisebox{-73pt}{\begin{tikzpicture}
\draw [color=blue!70!green, thick, decoration = {markings, mark=at position 0.5 with {\arrow[scale=1]{stealth}}}, postaction=decorate] (2.5,2) -- (2.5,3) node[right]{$a$} -- (2.5,3.75);
\draw [fill=black] (2.5,2) circle (0.04) node [below] {$\cO$};
\draw(2,0) -- (2,3) -- (3,4) -- (3,1) -- cycle;
\draw (2,0) node[below]{$Q$};
\end{tikzpicture}}
    \caption{SymTFT setup on $\Sigma \times I$, where $\Sigma$ is a $2$-manifold and $I$ is the interval. $\D$ is the reference Dirichlet boundary condition, and $\mathbb{B}_Q$ is the physical boundary condition. $V_\mu$ is the space of local operators which tells us how  the bulk anyons $\mu$ can end on the physical boundary.  $W_a^\mu$ is the space of junctions between $\mu$ and the symmetry line $a$ which lives on the reference boundary. Compactifying the interval leads to a $\cC$-symmetric 1+1d system $Q$, with the anyon $\mu$ turning into an $a$-twisted sector local operator $O$. 
    \cite{Lin_2023}
    }
    \label{fig:symTFT}
\end{figure}
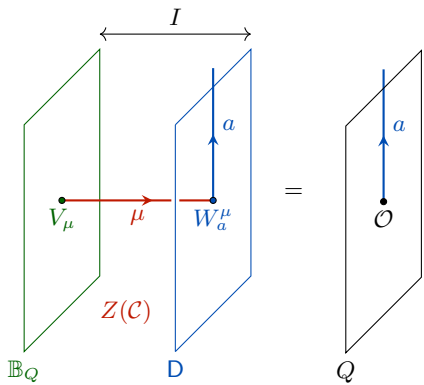

A generalized zero-form symmetry in $(1+1)$ dimensions may be described by a fusion category $\mathcal C$. On the lattice, the simple objects $a \in \rm{Irr} (\mathcal  C)$ are represented by matrix product operators (MPOs) $S_a$, whose multiplication reproduces the fusion rules of $\mathcal C$ \cite{Garre_Rubio_2023}. Ordinary finite-group symmetry is recovered as the special case $\mathrm{Vec}_G$, while categories such as $\mathrm{Rep}(G)$ provide basic examples of non-invertible symmetry.

A useful way to organize systems with categorical symmetry is through the symmetry topological field theory (SymTFT) \cite{Gaiotto_2021, bhardwaj2023generalized, moradi2023topological}. For a fusion-category symmetry $\mathcal C$, the associated $(2+1)$-dimensional topological quantum field theory (TQFT) is given by its Drinfeld center   $ Z(\mathcal C).$
The $(1+1)$-dimensional system may be viewed as a compactification of this bulk topological order on a slab; see Fig.~\ref{fig:symTFT}. One side of the slab is a distinguished gapped \emph{symmetry boundary}, which specifies the realization of the categorical symmetry, while the opposite, physical boundary specifies the gapped phase of the $(1+1)$-dimensional system.

Gapped boundaries of $Z(\mathcal C)$ are characterized by Lagrangian algebras. Equivalently, in the fusion-category language, they may be described by module categories over $\mathcal C$. Thus, given a fixed gapped boundary condition on the symmetry boundary,  a gapped phase can be encoded either by a module category over $\cC$ or by the associated Lagrangian algebra in $Z(\mathcal C)$ characterizing the physical boundary condition. Ref.~\cite{aksoy2025} formulates the classification of gapped phases directly in terms of Lagrangian algebras of the symTFT.

For the present work we are interested in symmetry-preserving phases with a unique ground state, which we refer to as SPT phases. Mathematically, these correspond to magnetic Lagrangian algebras or, equivalently, module categories with a single simple object \cite{thorngren2024fusion,zhang2024anomalies, inamura202411dsptphasesfusion}. A key feature of non-invertible symmetry is that distinct SPT phases need not all be related in the same way. Unlike ordinary SPT phases,  which can be connected by symmetry-preserving braided autoequivalences of $Z(\cC)$, non-invertible SPT phases can lie in different classes depending on the type of  braided autoequivalences required to connect them.

\subsection{Fixed-charge and fixed-algebra dualities}

Ref.~\cite{aksoy2025} organizes these relations using braided autoequivalences of the symTFT. A braided autoequivalence
\begin{equation}
    \cD : Z(\mathcal C)\longrightarrow Z(\mathcal C)
\end{equation}
permutes the anyon sectors while preserving their fusion and braiding data. The distinction relevant here is how $\cD$ acts on the Lagrangian algebra $\mathcal A_{\rm sym}$ defining the symmetry boundary.

The most restrictive class is a \emph{fixed-charge duality} (FCD). An FCD fixes each simple anyon condensed at the symmetry boundary, i.e. 
$    \cD_{\rm FCD}(\mu)= \mu$ for all anyons $\mu$ which appear as summands of $\mathcal{A}_{\rm sym}$ (such anyons are \emph{condensed} on the symmetry boundary).
Microscopically, this is the categorical analogue of a globally symmetric entangler: the circuit commutes with the symmetry MPOs and hence leaves their charge labels fixed. For ordinary group symmetries, familiar SPT entanglers fall into this class.

A \emph{fixed-algebra duality} (FAD) relaxes this condition. It preserves the symmetry boundary as a condensable algebra,
\begin{equation}
    \cD_{\rm FAD}(\mathcal A_{\rm sym})
    \simeq
    \mathcal A_{\rm sym},
\end{equation}
but may permute its simple summands. In lattice language, the corresponding transformation need not commute with each symmetry MPO individually. Instead, it may permute the symmetry lines while preserving their fusion structure. An FCD is therefore the special case of an FAD with trivial charge permutation.

More general braided autoequivalences need  not preserve the symmetry charges, but can still keep the  symmetry categories equivalent; these  are called fixed-symmetry dualities (FSD). 

This distinction is already nontrivial for $\mathrm{Rep}(D_8)$. The three SPT phases belong to distinct FCD classes, consistent with the absence of symmetry-commuting entanglers between them~\cite{seifnashri2024cluster}. Nevertheless, Ref.~\cite{aksoy2025} finds an $S_3$ group of FADs that permutes the three phases. This is the categorical motivation for the symmetry-permuting entanglers constructed below.

\subsection{Distinction from lattice dualities from generalized gauging}

Categorical dualities of one-dimensional lattice models have also been developed systematically from a different, complementary perspective by Lootens, Delcamp, Verstraete, and collaborators~\cite{Lootens_2023,Lootens_2024,Lootens_2025}. In that framework, realizations of a symmetric bond algebra are classified by module categories, and dualities are constructed between different module-category realizations. MPO intertwiners built from module-functor data then map the corresponding lattice models and their topological sectors.

This construction is closely related to the symTFT picture, but it addresses a different notion of duality from the one considered here. Changing the module category changes the realization of the symmetry boundary and is naturally interpreted as a generalized gauging, or a change of symmetry boundary, which we will refer to as  \emph{passive dualities}. Such dualities can exchange symmetry-preserving and symmetry-breaking phases and, in general, map local order operators to nonlocal string-order operators. Correspondingly, their general unitary realizations need not be finite-depth local circuits; the constructions of Ref.~\cite{Lootens_2025}, for example, have linear depth in general. 

By contrast, the FCDs and FADs relevant here keep the symmetry boundary fixed and act on the physical boundary -- in contradistinction to passive dualities, these may be called \emph{active dualities}. Since these active dualities can sometimes map one SPT phase  to another, it is natural to ask whether they admit a  microscopic realization by an FDLU acting on a fixed Hilbert space. This is the question addressed in the present work.

\subsection{Microscopic realization by quantum circuits}

When the symmetries form a finite abelian group $G$, the relation between duality automorphisms and microscopic locality-preserving transformations is understood much more systematically. Ref.~\cite{Ma2026quantumcellular} classifies quantum cellular automata (QCA) acting on the symmetric local operator algebra and constructs a surjective map from such QCAs to anyon permutations of the corresponding $(2+1)$-dimensional $G$-gauge theory.  They  reproduce microscopically the classification of braided autoequivalences for the quantum double of $G$ in e.g. Ref.~\cite{moradi2023topological}, Appendix B, including the special case of symmetric entanglers when the QCA can be extended to the full algebra of operators and potentially (if the QCA index is trivial) be realized by an FDLU.

For non-invertible symmetries,  Refs. ~\cite{Jones2024DHR, JonesSchatzWilliamson2026} studied the relation between QCAs on the symmetric subalgebra and braided autoequivalences of $Z(
\cC)$, and the conditions under which the QCA can be extended  to the full algebra, and hence potentially an FDLU. However, their focus was not on the construction of such FDLUs.
 
An explicit FDLU globally symmetric under non-invertible symmetry has been constructed in Ref.~\cite{You_2026}, for the example of $\cC = \Rep(A_4)$. This was inspired by the existence of an FCD for $\Rep(A_4)$ \cite{aksoy2025}. That example suggested that FCDs may admit microscopic representatives as globally symmetric entanglers, and overturned the proposed general inference from the absence of stacking structure that non-invertible SPT phases cannot admit symmetric entanglers  \cite{seifnashri2024cluster}. However, the explicit MPU was obtained essentially ad hoc, and no general procedure for constructing a microscopic circuit from the categorical duality data was given.

The present work addresses the next case in this hierarchy. For $\mathrm{Rep}(D_8)$, the relevant SPT phases are not connected by FCDs (in fact there are no FCDs for $\Rep(D_8)$ symmetry) and therefore cannot be related by symmetric entanglers. They are, however, related by FADs. We therefore seek finite-depth circuits $U$ that do not commute with the symmetry MPOs individually, but instead permute them. We call such circuits \emph{symmetry-permuting entanglers}. The remainder of the paper constructs an explicit $\mathrm{Rep}(D_8)$ example and develops a  way to extract its induced anyon permutation directly from microscopic string operators.

\section{Symmetry-permuting entanglers  from local crossing relations}
\label{sec:symmetry-permuting-entanglers}

We now turn to the microscopic realization of an FAD as an MPU.   
We will first derive conditions that the local MPO tensors of the FDLU $U$ and the symmetry MPO $S_a$ need to satisfy. We start from the global symmetry-permuting condition
\begin{equation}
U S_a U^\dagger
=
S_{\varphi(a)},
\qquad a \in \rm{Irr} (\mathcal C) ,
\label{eq:global-symmetry-permutation}
\end{equation}
for all sufficiently large periodic chains. Here, $\varphi$ is a permutation of the simple objects compatible with fusion. We refer to such a circuit as a \emph{symmetry-permuting entangler}. When $\varphi=\mathrm{id}$, this reduces to a symmetric entangler.


\subsection{From global symmetry permutation to local pulling-through}
\label{subsec:global-to-local}

It is useful to regard an MPO as an MPS by vectorizing its physical input and output indices. Under vectorization,
\begin{equation}
|U S_a\rangle\rangle
=
(U\otimes \mathds{1})|S_a\rangle\rangle ,
\end{equation}
whereas
\begin{equation}
|S_{\varphi(a)} U\rangle\rangle
=
(\mathds{1}\otimes U^T)
|S_{\varphi(a)}\rangle\rangle .
\end{equation}
The symmetry-permuting relation therefore becomes an equality of two MPS families on the doubled Hilbert space,
\begin{equation}
(U\otimes\mathds{1})|S_a\rangle\rangle
=
(\mathds{1}\otimes U^T)|S_{\varphi(a)}\rangle\rangle .
\label{eq:vectorized-global-relation}
\end{equation}

We take $S_a$ to be a minimal tensor for the simple symmetry sector $a$. After a finite amount of blocking, its vectorization can be chosen normal. 

A normal MPS acted upon by a  finite-depth circuit still admits a normal minimal MPS representation. Note that directly contracting the circuit with the MPS may enlarge the raw virtual space and introduce redundant virtual directions, so the resulting unreduced tensor need not itself be normal. We therefore pass to the \emph{minimal canonical representations} of the two sides of Eq.~\eqref{eq:vectorized-global-relation}.

Let $A_a$ and $B_a$ denote minimal normal MPS tensors representing respectively
\begin{equation}
(U\otimes1)|S_a\rangle\rangle ,
\qquad
(1\otimes U^T)|S_{\varphi(a)}\rangle\rangle .
\end{equation}
Because their periodic MPS families coincide for all sufficiently large lengths, after a fixed amount of blocking if necessary the fundamental theorem of matrix product states gives an invertible virtual intertwiner $\mathcal X_a$ satisfying
\begin{equation}
A_a^I\,\mathcal X_a
=
\mathcal X_a\,B_a^I .
\label{eq:general-bulk-zipper}
\end{equation}
Here $I$ denotes a possibly blocked physical index.

\begin{figure*}[t]
\centering
\begin{tikzpicture}[scale=1.0,baseline={(current bounding box.center)}]

\tikzset{
  phys/.style={black, very thick},
  symvirt/.style={
    blue!70!black,
    very thick,
    decorate,
    decoration={snake, amplitude=0.55mm, segment length=3.0mm}
  },
  entvirt/.style={red!75!black, very thick},
  symtensor/.style={
    circle,
    draw=black,
    very thick,
    minimum size=8mm,
    inner sep=0pt,
    fill=white
  },
  enttensor/.style={
    rectangle,
    draw=black,
    very thick,
    minimum width=8mm,
    minimum height=8mm,
    inner sep=0pt,
    fill=white
  },
  zipper/.style={
    ellipse,
    draw=black,
    very thick,
    minimum width=7mm,
    minimum height=10mm,
    inner sep=0pt,
    fill=white
  }
}


\node[symtensor] (symL) at (0,1.6) {$a$};
\node[enttensor] (entL) at (0,0) {$$};

\draw[phys] (0,2.75) -- (symL.north);
\draw[phys] (symL.south) -- (entL.north);
\draw[phys] (entL.south) -- (0,-1.15);

\draw[symvirt] (-1.35,1.6) -- (symL.west);
\draw[symvirt] (symL.east) -- (1.35,1.6);

\draw[entvirt] (-1.35,0) -- (entL.west);
\draw[entvirt] (entL.east) -- (1.35,0);

\node at (2.35,0.8) {\Large $=$};


\node[enttensor] (entR) at (5.7,1.6) {$ $};
\node[symtensor] (symR) at (5.7,0) {$\varphi(a)$};

\draw[phys] (5.7,2.75) -- (entR.north);
\draw[phys] (entR.south) -- (symR.north);
\draw[phys] (symR.south) -- (5.7,-1.15);

\node[zipper] (zipL) at (4.0,0.8) {$\mathcal X_a$};
\node[zipper] (zipR) at (7.4,0.8) {$\mathcal X_a^{-1}$};

\coordinate (zipLNW) at ($(zipL.west)+(0,0.18)$);
\coordinate (zipLSW) at ($(zipL.west)+(0,-0.18)$);
\coordinate (zipLNE) at ($(zipL.east)+(0,0.18)$);
\coordinate (zipLSE) at ($(zipL.east)+(0,-0.18)$);

\coordinate (zipRNW) at ($(zipR.west)+(0,0.18)$);
\coordinate (zipRSW) at ($(zipR.west)+(0,-0.18)$);
\coordinate (zipRNE) at ($(zipR.east)+(0,0.18)$);
\coordinate (zipRSE) at ($(zipR.east)+(0,-0.18)$);


\draw[symvirt] (2.95,1.6) -- (zipLNW);
\draw[entvirt] (2.95,0.0) -- (zipLSW);

\draw[entvirt] (zipLNE) -- (entR.west);
\draw[symvirt] (zipLSE) -- (symR.west);


\draw[entvirt] (entR.east) -- (zipRNW);
\draw[symvirt] (symR.east) -- (zipRSW);

\draw[symvirt] (zipRNE) -- (8.45,1.6);
\draw[entvirt] (zipRSE) -- (8.45,0.0);

\end{tikzpicture}

\caption{
Diagrammatic form of the local pulling-through equation. Black lines denote physical legs.
The symmetry MPO tensor is drawn as a circle with blue  virtual legs, the entangler
MPO tensor as a square with red  virtual legs, and the crossing tensor $\mathcal{X}_a$  implements the exchange of the symmetry and entangler
virtual spaces.
}
\label{fig:bulk-pulling-through}
\end{figure*}
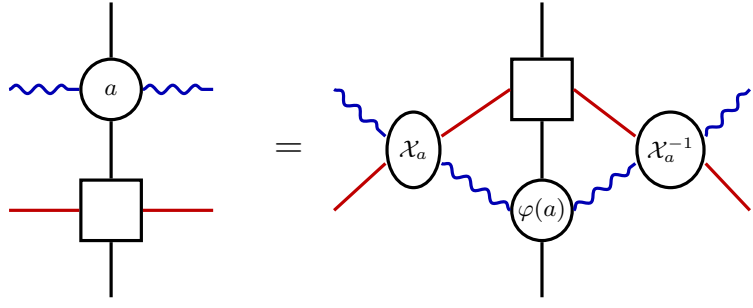

If the raw contracted tensors are already normal,  this equation becomes 
\begin{equation}
   \sum_{j} T_{S_a}^{i, j}  \otimes M^{j, k}= \mathcal{X}_a \left(\sum_j M^{i,j} \otimes T_{S_a}^{j,k}   \right) \mathcal{X}_a^{-1}, 
   \label{eq:pulling_through}
\end{equation}
 where we denote the MPO tensors of $U$ by $M$ and the MPO tensors of $S_a$ by $T_{S_a}$. This is expressed diagrammatically by Fig. \ref{fig:bulk-pulling-through}. We call $\mathcal{X}_a$ the \emph{crossing tensor} for symmetry $S_a$.

\subsection{\texorpdfstring{$\operatorname{Rep}(D_8)$}{Rep(D8)} fixed points and the target FAD}
\label{subsec:repD8-target-FAD}

We now specialize to
\begin{equation}
D_8=
\langle r,s\mid r^4=s^2=1,\;srs=r^{-1}\rangle ,
\end{equation}
and define
\begin{equation}
H=\langle u=r^2,v=s\rangle\simeq\mathbb Z_2\times\mathbb Z_2 .
\label{eq:H-subgroup}
\end{equation}
We order the physical group basis as
\begin{equation}
(e,u,v,uv\mid r,ru,rv,ruv).
\label{eq:physical-basis-order}
\end{equation}

The simple objects of $\operatorname{Rep}(D_8)$ are the four one-dimensional representations
\begin{equation}
\mathbf 1,\quad \rho_a,\quad \rho_b,\quad \rho_c=\rho_a\rho_b,
\end{equation}
together with the two-dimensional representation $\sigma$. We choose conventions
\begin{equation}
\rho_a(r)=1,\qquad \rho_a(s)=-1,
\end{equation}
\begin{equation}
\rho_b(r)=-1,\qquad \rho_b(s)=1,
\end{equation}
and
\begin{equation}
\sigma(r)=-iY,\qquad \sigma(s)=Z.
\label{eq:sigma-convention}
\end{equation}

The particular FAD that we wish to realize is the involution appearing in the third line of Eq.~(C7) of Ref.~\cite{aksoy2025}, namely the one containing the transposition $(i,m)$. In their notation its full action is
\begin{equation}
(i,m)(n,k)(u,r)(v,t)(b,c)(g,e).
\label{eq:aksoy-wen-third-line}
\end{equation}
As detailed in Appendix~\ref{app:anyon-dictionary}, translating this permutation into conjugacy-class and centralizer-representation labels shows that it exchanges the trivial SPT with the SPT associated with
\begin{equation}
H=\langle r^2,s\rangle .
\end{equation}
On the symmetry charges themselves, its action is
\begin{equation}
\rho_a\longleftrightarrow \rho_c,
\qquad
\rho_b\longmapsto \rho_b,
\qquad
\sigma\longmapsto\sigma .
\label{eq:target-symmetry-permutation}
\end{equation}

The existence of this FAD is the categorical reason to expect that a finite-depth entangler may exist despite the obstruction to a symmetry-commuting circuit. Rather than searching for an arbitrary MPU, we therefore seek a microscopic circuit realizing precisely the symmetry permutation~\eqref{eq:target-symmetry-permutation}.

We note that there is another FAD which can be realized as an on-site unitary circuit. Consider a representative of the nontrivial outer group automorphism class  given by \footnote{The choice $f(r) = r$  differs from this one by an inner automorphism and does not affect the corresponding FAD.}
\begin{equation}
     f(r)=r^{-1},\qquad f(s)=rs, 
\end{equation}
which acts on the on-site Hilbert space by the corresponding automorphism:
\begin{equation}
U_f=\bigotimes_j U_{f,j},\quad U_{f, j} \ket{g}_j =\ket{f(g)}_j.
\label{eq:onsite_FAD}
\end{equation}
This induces a corresponding permutation of anyons, which corresponds to 
\begin{equation}
(m,j)(l,n)(u,s)(q,v)(a,b)(e,h)
 \label{eq:onsite_FAD_anyon}
\end{equation}
in the notation of Ref. \cite{aksoy2025}, and exchanges the two nontrivial SPT phases. 

The main task is therefore to construct a second, non-on-site generator connecting the trivial phase to one of the nontrivial SPTs, such that combined with the above on-site unitary, we generate the full $S_3$ group of FADs.\footnote{Ref.~\cite{meng2025noninvertiblesptsonsiterealization} discusses an $S_3$ duality relating the three $\Rep(D_8)$ SPT phases, but their duality involves modifying the Hilbert space and is not manifestly a finite-depth circuit.}

We construct this second generator which maps between the trivial product-state fixed point and the $H$-based SPT state. The two MPS are given in Ref.~\cite{meng2025noninvertiblesptsonsiterealization}. The product state MPS is
\begin{equation}
A_0^g=\delta_{g,e},
\label{eq:product-MPS}
\end{equation}
while the target SPT phase MPS, which comes from the module category of $\Rep^\omega(H)$ where $\omega$ is the nontrivial 2-cocycle of $H$, has bond dimension two and is defined by 
\begin{equation}
A_\omega^g=\frac12Q_g,
\qquad g\in H,
\label{eq:H-SPT-MPS}
\end{equation}
with $A_\omega^g=0$ for $g\notin H$, where
\begin{equation}
Q_e=I,\qquad
Q_u=-X,\qquad
Q_v=-Z,\qquad
Q_{uv}=iY .
\label{eq:Q-matrices}
\end{equation}

\subsection{Constraints from crossing tensors}
\label{subsec:action-to-zipper}
We make the bond-dimension-two ansatz that Eq.~\eqref{eq:pulling_through} holds directly for the MPU tensors. Although motivated by the global pulling-through relation, the consistency of this ansatz is not guaranteed a priori; below we show that it admits a unitary solution.

Because the MPU is required to map one SPT fixed point to the other, the corresponding MPS action tensors provide additional constraints that determine the crossing tensors $\mathcal X_a$.\footnote{In general, the circuit should be constructible from the data of the braided autoequivalence itself without reference to the phases it connects; here the fixed-point MPS data provide useful additional input.}

 For a symmetry line $S_a$, let
\begin{equation}
\Lambda_a^\mu:V_a\otimes V_D \longrightarrow V_D
\label{eq:MPS-action-tensor}
\end{equation}
denote the tensor implementing the action of the symmetry MPO (with virtual space $V_a$) on the virtual space $V_D$ of an MPS \cite{Garre_Rubio_2023}. The product state and the $H$-SPT carry different action tensors,
\begin{equation}
\Lambda_a^{(0)},
\qquad
\Lambda_a^{(\omega)}.
\end{equation}

When an MPU maps the product MPS to the SPT MPS, its bond space becomes part of the virtual space of the output state. Compatibility of the symmetry action on the two sides therefore requires an intertwiner $\mathcal X_a$ relating the two sets of action tensors, and this allows us to obtain the crossing tensors  from action tensors -- we refer to Appendix~\ref{app:action_tensors} for details.

For the present pair of fixed points, a convenient gauge gives us the following crossing tensors for invertible symmetry lines:
\begin{equation}
\mathcal X_{\mathbf 1}=I,\qquad
\mathcal X_{\rho_b}=I,\qquad
\mathcal X_{\rho_a}=X,\qquad
\mathcal X_{\rho_c}=X,
\label{eq:invertible-zippers}
\end{equation}
while for the non-invertible line,
\begin{equation}
\mathcal X_\sigma
=
|0\rangle\langle0|\otimes Z
+
|1\rangle\langle1|\otimes Y .
\label{eq:sigma-zipper}
\end{equation}
(Here, $
|0\rangle\langle0|$ and $
|1\rangle\langle1| $ act on the  symmetry MPO virtual space  while $Z$ and $Y$ act on the entangler MPO virtual  space.)

The explicit rectangular action tensors $\Lambda_a^{(0)}$ and $\Lambda_a^{(\omega)}$ from which we derive Eqs.~\eqref{eq:invertible-zippers}--\eqref{eq:sigma-zipper} are given in Appendix~\ref{app:action_tensors}. 

Let
\begin{equation}
M=
\sum_{g,h\in D_8}
|g\rangle\langle h|\otimes M^{g,h}
\label{eq:local-MPU-tensor}
\end{equation}
be the putative bond-dimension-two MPU tensor. Using the crossing tensors (Eqs. ~\eqref{eq:invertible-zippers} and \eqref{eq:sigma-zipper}), we get the following local crossing relations from Eq. \eqref{eq:pulling_through}:
\begin{equation}
\varphi(\chi)(g)\,
M^{g,h}\mathcal X_\chi
=
\chi(h)\,
\mathcal X_\chi M^{g,h},
\qquad
\chi=\mathbf 1,\rho_a,\rho_b,\rho_c ,
\label{eq:invertible-pulling-through}
\end{equation}
and
\begin{equation}
[\sigma(g)\otimes M^{g,h}]
\mathcal X_\sigma
=
\mathcal X_\sigma
[\sigma(h)\otimes M^{g,h}] .
\label{eq:sigma-pulling-through}
\end{equation}

The $\rho_b$-equation already fixes the basic block structure. Because
\begin{equation}
\varphi(\rho_b)=\rho_b,\qquad \mathcal X_{\rho_b}=I,
\end{equation}
a nonzero tensor $M^{g,h}$ requires $\rho_b(g)=\rho_b(h).$
Hence
\begin{equation}
M^{g,h}=0
\qquad
\text{unless}\qquad
g,h\in H
\quad\text{or}\quad
g,h\in rH,
\label{eq:block-selection}
\end{equation}
so the local tensor decomposes into two physical blocks,
\begin{equation}
M=
\begin{pmatrix}
M_H&0\\
0&M_{rH}
\end{pmatrix}.
\label{eq:E-block-form}
\end{equation}

Solving Eqs.~\eqref{eq:invertible-pulling-through}--\eqref{eq:sigma-pulling-through} then fixes every nonzero $2\times2$ block up to a scalar factor.  Explicitly, one finds
\begin{equation}
M^{h,k}=\alpha_{h,k}Q_{hk},
\qquad
M^{rh,rk}=\beta_{h,k} K_{hk}.
\label{eq:scalar-reduction}
\end{equation}
for some coefficients $\alpha_{h,k}$ and $\beta_{h,k}$, where 
\begin{align}
K_e=-Z-Y,\qquad
K_{u}=-Z+Y, \nonumber \\ 
K_v=I+iX,\qquad
K_{uv}=I-iX.
\label{eq:appB-R-matrices}
\end{align}

Note that if one instead imposes $\varphi=\mathrm{id}$, the pulling-through equations eliminate the nonzero $rH$ block and thus spoil unitarity, reproducing  the obstruction to a symmetric entangler. \footnote{However, as in Ref.~\cite{seifnashri2024cluster}, this shows the nonexistence of a symmetric entangler for a  particular model realizing the $\Rep(D_8)$ SPT phases, rather than being a general topological obstruction. 
}

\subsection{The symmetry-permuting MPU}
\label{subsec:final-MPU}

The remaining scalar freedom can be  fixed by physical requirements. 
First, the MPU must map the chosen product fixed point exactly to the chosen SPT fixed point,
\begin{equation}
U|\Psi_0\rangle=|\Psi_\omega\rangle .
\label{eq:exact-state-mapping}
\end{equation}
which immediately fixes 
\begin{equation}
M^{g,e} = A_\omega^g.
\end{equation}
We then impose unitarity ($UU^\dagger = \mathds{1}$) and also demand that the circuit satisfies $U^2 = \mathds{1}$ since the FAD which we wish to realize is order-2. A convenient form of the unitarity condition is given by Ref.~\cite{shukla2025simplegeneralequationmatrix}. Once unitarity is assumed, $U^2 = \mathds{1}$ is equivalent to Hermiticity. 
 On the individual tensors, Hermiticity can be imposed up to a fixed virtual gauge transformation; a choice which is consistent with our matrices is:
\begin{equation}
\overline{M^{h,g}}
=
Z M^{g,h} Z.
\label{eq:local-hermiticity}
\end{equation}

After imposing these conditions, we obtain the final result: 
\begin{align}
M^{h,k}
&=
\frac{(-1)^{m'}}{2}Q_{hk}, \nonumber\\
M^{rh,rk}
&=
\frac{(-1)^{(m+n)(m'+n')}}{2\sqrt2}
K_{hk},
\label{eq:final}
\end{align}
where $m, m', n, n' \in\{0,1\}$ are defined by  $$ h=u^mv^n,
\qquad
k=u^{m'}v^{n'}.$$

MPUs are QCAs, and QCAs with trivial index are FDLUs \cite{Ignacio_Cirac_2017}. Our entangler is an MPU that squares to the identity independent of system size, and since the index is additive under composition of MPUs, we have $$\rm{ind} (U^2)  = 2 \rm{ind} (U) = \rm{ind} (\mathds{1}) = 0,$$ 
so $\rm{ind}(U) = 0$, and the MPU lies in the FDLU class.


\begin{figure*}[t]
\centering
\begin{tikzpicture}[x=1cm,y=1cm,baseline={(current bounding box.center)}, scale = 0.9]

\tikzset{
  phys/.style={black, very thick},
  symvirt/.style={
    blue!70!black,
    very thick,
    decorate,
    decoration={snake, amplitude=0.5mm, segment length=3mm}
  },
  entvirt/.style={red!75!black, very thick},
  symtensor/.style={
    circle,
    draw=black,
    very thick,
    minimum size=8mm,
    inner sep=0pt,
    fill=white
  },
  enttensor/.style={
    rectangle,
    draw=black,
    very thick,
    minimum width=8mm,
    minimum height=8mm,
    inner sep=0pt,
    fill=white
  },
  endpoint/.style={
    diamond,
    aspect=1.15,
    draw=black,
    very thick,
    minimum width=7.5mm,
    minimum height=7.5mm,
    inner sep=0.5pt,
    fill=white
  },
  zipper/.style={
    ellipse,
    draw=black,
    very thick,
    minimum width=8mm,
    minimum height=10mm,
    inner sep=0.5pt,
    fill=white
  },
  stagearrow/.style={
    -{Latex[length=2.5mm]},
    thick
  }
}


\node[endpoint]  (psiA) at (0,1.2) {$\Gamma_a$};
\node[symtensor] (symA) at (2.7,1.2) {$a$};

\node[enttensor] (entA1) at (0,-1.2) {$ $};
\node[enttensor] (entA2) at (2.7,-1.2) {$ $};

\draw[phys] (0,2.25) -- (psiA.north);
\draw[phys] (psiA.south) -- (entA1.north);
\draw[phys] (entA1.south) -- (0,-2.25);

\draw[phys] (2.7,2.25) -- (symA.north);
\draw[phys] (symA.south) -- (entA2.north);
\draw[phys] (entA2.south) -- (2.7,-2.25);

\draw[symvirt] (psiA.east) -- (symA.west);
\draw[symvirt] (symA.east) -- (3.85,1.2);

\draw[entvirt] (-1.05,-1.2) -- (entA1.west);
\draw[entvirt] (entA1.east) -- (entA2.west);
\draw[entvirt] (entA2.east) -- (3.85,-1.2);

\draw[stagearrow] (4.8,0) -- (6.5,0)
  node[midway,above=3pt,align=center] {\small bulk\\[-1pt]\small pull-through};


\node[endpoint]  (psiB)  at (7.7,1.2) {$\Gamma_a$};
\node[enttensor] (entB2) at (10.4,1.2) {$ $};

\node[enttensor] (entB1) at (7.7,-1.2) {$ $};
\node[symtensor] (symB)  at (10.4,-1.2) {$\varphi(a)$};

\node[zipper]    (zipB)  at (9.05,0) {$\mathcal X_a$};

\draw[phys] (7.7,2.25) -- (psiB.north);
\draw[phys] (psiB.south) -- (entB1.north);
\draw[phys] (entB1.south) -- (7.7,-2.25);

\draw[phys] (10.4,2.25) -- (entB2.north);
\draw[phys] (entB2.south) -- (symB.north);
\draw[phys] (symB.south) -- (10.4,-2.25);

\draw[entvirt] (6.65,-1.2) -- (entB1.west);
\draw[entvirt] (entB2.east) -- (11.55,1.2);
\draw[symvirt] (symB.east) -- (11.55,-1.2);

\coordinate (zipBLu) at ($(zipB.west)+(0,0.18)$);
\coordinate (zipBLd) at ($(zipB.west)+(0,-0.18)$);
\coordinate (zipBRu) at ($(zipB.east)+(0,0.18)$);
\coordinate (zipBRd) at ($(zipB.east)+(0,-0.18)$);

\draw[symvirt]
  (psiB.east) to[out=0,in=170] (zipBLu);
\draw[symvirt]
  (zipBRd) to[out=-10,in=180] (symB.west);

\draw[entvirt]
  (entB1.east) to[out=0,in=190] (zipBLd);
\draw[entvirt]
  (zipBRu) to[out=10,in=180] (entB2.west);


\draw[stagearrow] (12.5,0) -- (14.2,0)
  node[midway,above=3pt,align=center] {\small endpoint\\ pull-through\small};


\node[enttensor] (entC1) at (15.4,1.2) {$ $};
\node[enttensor] (entC2) at (18.1,1.2) {$ $};

\node[endpoint]  (psiC)  at (15.4,-1.2) {$\widetilde\Gamma_{\varphi(a)}$};
\node[symtensor] (symC)  at (18.1,-1.2) {$\varphi(a)$};

\draw[phys] (15.4,2.25) -- (entC1.north);
\draw[phys] (entC1.south) -- (psiC.north);
\draw[phys] (psiC.south) -- (15.4,-2.25);

\draw[phys] (18.1,2.25) -- (entC2.north);
\draw[phys] (entC2.south) -- (symC.north);
\draw[phys] (symC.south) -- (18.1,-2.25);

\draw[symvirt] (psiC.east) -- (symC.west);
\draw[symvirt] (symC.east) -- (19.25,-1.2);

\draw[entvirt] (14.35,1.2) -- (entC1.west);
\draw[entvirt] (entC1.east) -- (entC2.west);
\draw[entvirt] (entC2.east) -- (19.25,1.2);

\end{tikzpicture}

\caption{
Endpoint pulling-through in three stages. 
Left: the initial local configuration with endpoint tensor $\Gamma_a$ above the entangler.
Middle: after applying bulk pulling-through, the endpoint is still the original $\Gamma_a$ on the top, while the symmetry line has moved to the bottom and the residual crossing tensor $\mathcal X_a$ explicitly connects the top endpoint leg to the bottom network. 
Right: $\mathcal{X}_a$ is absorbed into a new endpoint tensor $\widetilde \Gamma_{\varphi(a)}$, yielding the transformed endpoint equation.
}
\label{fig:endpoint-pulling-through-three-stage}
\end{figure*}
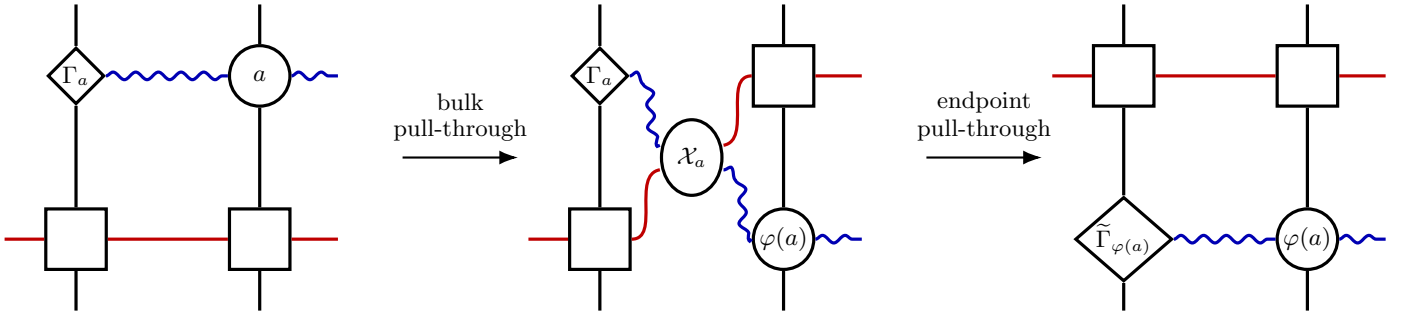

\section{Anyon permutation action of the  \texorpdfstring{$\operatorname{Rep}(D_8)$}{Rep(D8)} symmetry-permuting entangler}
\label{sec:anyon-permutations}
The construction above was designed to realize a particular FAD, but the symmetry permutation alone does not determine its action on  anyon sectors. We now show microscopically that the MPU transforms string operators according to the anyon permutation of the corresponding FAD.

In a 1+1D $\mathcal C$-symmetric system, an anyon of the symTFT $Z(\mathcal C)$ is represented by a string operator: a symmetry string terminated by charged endpoint operators. For non-invertible symmetry, the string itself is an MPO \cite{chavda2026algebrasorderparametersonedimensional}. 

For $\Rep(G)$ symmetry, $Z(\Rep(G)) \simeq \Vc_G^G$, the category of $G$-graded $G$-representations, and a simple object $W$ of $\Vc_G^G$ corresponds to an anyon.  Concretely, a multiplet of  string operators corresponding to $W$ can be constructed from $G$-equivariant maps 
$$\varphi^W_S: S \ra W$$
for irreps $S$ in $\Rep(G)$ and grade-preserving maps 
$$\Gamma: W \ra \End(\cH_i)$$ where $\cH_i$ is the local Hilbert space (which in our case is just $\CC[G]$ with the obvious grading).  $\Gamma$ caps off the symmetry MPO at the endpoint, and for each basis vector $w_{g, c} \in W$ (here, $g$ is the grading and $c$ is an additional degeneracy index for the same grading),   we get a physical on-site operator $\Gamma(w_{g,c})$ \cite{chavda2026algebrasorderparametersonedimensional}. 

We determine the action of the entangler $U$ on anyons via its action on the corresponding string operators. Concretely, pulling the bulk of the string through $U$  converts the bulk string from $S_a$ to $S_{\varphi(a)}$; neighboring crossing tensors cancel in the interior, leaving a single crossing tensor adjacent to each endpoint. As illustrated in Fig.~\ref{fig:endpoint-pulling-through-three-stage}, after the bulk pulling-through, the original endpoint tensor $\Gamma_a $ remains in place while the symmetry string has moved to the opposite side of the circuit. The remaining crossing tensor is then absorbed into the endpoint, producing a transformed endpoint tensor $\widetilde\Gamma_{\varphi(a)}$.
At the endpoint, the crossing equation  reads 
\begin{equation}\label{eq:endpoint}
 \left( \sum_n (\Gamma_a)_\alpha^{m,n} \otimes M^{n,l} \right) \mathcal{X}_a
 = 
 \sum_{n} M^{m,n} \otimes 
 (\widetilde \Gamma_{\varphi(a)})_\alpha^{n,l}
\end{equation}
where $\mathcal X_a$ is the same crossing tensor of Eq. \eqref{eq:general-bulk-zipper}, uncanceled at the endpoint, and $\widetilde\Gamma_{\varphi(a)}$ is the transformed endpoint operator after pulling the original endpoint through and absorbing the crossing tensor ($\alpha$ labels the component of the endpoint operator, while $m,n,l$ are physical space indices). This determines the transformed endpoint operator $\widetilde \Gamma_{\varphi(a)}$. 

We now apply this transformation to the string operators which gain a VEV in the trivial product state. 

\subsection{Invertible strings}

Consider the anyon $([e], \rho_b)$. The corresponding object $W$ is trivially graded with underlying $G$-representation $\rho_b$, so  the endpoint operator is diagonal. Since $\mathcal{X}_{\rho_b}$ is trivial (up to swapping the two factors), we obtain 
$$\widetilde\Gamma_{\rho_b} = \Gamma_{\rho_b}.$$ 
Thus,  we end up with the same anyon $([e], \rho_b)$. 

We then consider the anyon $([e], \rho_a)$. The corresponding object $W$ is again trivially graded, with underlying $G$-representation $\rho_a$, so we take the endpoint operator $\Gamma_{\rho_a}$ to be diagonal. Now, $\mathcal{X}_{\rho_a} = X$, which leads to
\begin{align}
    (\Gamma_{\rho_a} \otimes M^{g,h} ) X = M^{g,h}   \otimes \widetilde\Gamma_{\rho_c}.
\end{align}
Using the fact that 
\begin{equation}
     M^{g,h} X = \lambda_h M^{g, hr^2}, 
     \label{eq:X_relation}
\end{equation}
where $$\lambda_h = (1,1,1,1 | -i, i ,i, -i ),$$
we obtain 
$$\widetilde\Gamma_{\rho_c} = \Gamma_{\rho_a} \Lambda R_{r^2}$$
where $\Lambda$ is a diagonal matrix consisting of phases  $\lambda_h$ on the diagonals, and $R_{g}$ is the on-site operator implementing right-multiplication by $g$: $$R_g|h\rangle=|hg^{-1}\rangle.$$ 
Since both $\Gamma_{\rho_a}$ and $\Lambda$ are diagonal and have trivial grading, $\widetilde\Gamma_{\rho_c}$ has grading $r^2$ coming from $R_{r^2}$. 

Recall that the entangler transforms the bulk symmetry from $\rho_a$ to $\rho_c$. Hence the resulting anyon is $([r^2], \rho_c)$. Similarly, $([e], \rho_c)$ transforms into $([r^2], \rho_a)$.

\subsection{The noninvertible \texorpdfstring{$\sigma$}{sigma} string}
\label{subsec:sigma-string-anyon-map}

The most instructive example is the noninvertible symmetry line $\sigma$. Recall the crossing tensor
\begin{equation}
\mathcal X_\sigma
=
|0\rangle\langle0|\otimes Z
+
|1\rangle\langle1|\otimes Y .
\label{eq:anyons-sigma-zipper}
\end{equation}
The bulk symmetry label itself is fixed,
\begin{equation}
\sigma\longmapsto\sigma,
\label{eq:crossing_sigma}
\end{equation}
but the two components of the virtual symmetry leg are acted on differently by the crossing tensor $\mathcal{X}_\sigma$.

Consider the anyon  $([e],\sigma)$. The corresponding graded representation is  concentrated in the identity grade, with underlying $D_8$-irrep $\sigma$, thus the endpoint operator is diagonal on $\CC[G]$. A convenient choice  for the two components is 
\begin{align}
    (\Gamma_\sigma)_0 = D_H, \qquad (\Gamma_\sigma)_1 = i D_H
\end{align}
where 
$$D_H = \rm{diag}(1,-1,1,-1 | 0, 0, 0, 0).$$

$M^{g,h}$ satisfy 
\begin{align}
    M^{g,h} Z = -(-1)^{m +m'} M^{g, hs}, \nonumber\\
  M^{g,h} Y =  i (-1)^{m+m'}  M^{g, hr^2s} .
\end{align}
Thus, from
\begin{align}
   \left( (\Gamma_\sigma)_0^g \otimes M^{g,h} \right) Z = \sum_{k} M^{g, k } \otimes (\widetilde\Gamma_\sigma)_0^{k,h}  
\end{align}
 we obtain $$(\widetilde\Gamma_\sigma)_0 =  -R_s D_H,$$ and similarly, we get $$(\widetilde\Gamma_\sigma)_1 =- R_{r^2s} D_H.$$
The transformed endpoint tensors therefore carry gradings
\begin{equation}
|0\rangle_\sigma\leftrightarrow s,
\qquad
|1\rangle_\sigma\leftrightarrow r^2s.
\label{eq:sigma-transformed-gradings}
\end{equation}
These two grades form the conjugacy class
\begin{equation}
[s]=\{s,r^2s\}.
\label{eq:s-conjugacy-class}
\end{equation}

To determine the full anyon label, we restrict the irrep $\sigma$ to the centralizer
\begin{equation}
Z_{D_8}(s)=\langle r^2,s\rangle.
\label{eq:s-centralizer}
\end{equation}
With our conventions,
\begin{equation}
\sigma(r^2)=-I,
\qquad
\sigma(s)=Z.
\label{eq:sigma-centralizer-action}
\end{equation}
On the $s$-graded component, corresponding to $|0\rangle_\sigma$,
\begin{equation}
\sigma(r^2)|0\rangle=-|0\rangle,
\qquad
\sigma(s)|0\rangle=|0\rangle.
\label{eq:sigma-on-s-graded-component}
\end{equation}
Hence the transformed endpoint carries the one-dimensional centralizer representation $\eta$ of $H$ such that
\begin{equation}
\eta(r^2)=-1,
\qquad
\eta(s)=+1.
\label{eq:eta-definition}
\end{equation}
This is precisely the irrep $\omega_2$ in the dictionary of  Appendix \ref{app:anyon-dictionary}. Thus,
\begin{equation}
([e],\sigma)
\longmapsto
([s],\omega_2).
\label{eq:sigma-anyon-map}
\end{equation}

\subsection{Anyon action of the \texorpdfstring{$\operatorname{Rep}(D_8)$}{Rep(D8)} MPU}
\label{subsec:repD8-anyon-action}
Collecting these results,  the action on the anyons condensed at the trivial SPT boundary is
\begin{equation}
\boxed{
\begin{aligned}
([e],\mathbf 1)&\longmapsto([e],\mathbf 1),\\
([e],\rho_b)&\longmapsto([e],\rho_b),\\
([e],\rho_a)&\longmapsto([r^2],\rho_c),\\
([e],\rho_c)&\longmapsto([r^2],\rho_a),\\
([e],\sigma)&\longmapsto([s],\omega_2).
\end{aligned}}
\label{eq:condensed-sector-action}
\end{equation}
Comparison with the \(\Rep(D_8)\) FAD classification of Ref.~\cite{aksoy2025} uniquely identifies the FAD realized by our circuit. The same computation can in principle be carried out for all 22 anyons. The only additional subtlety arises when $W$ is irreducible as a graded representation but reducible after forgetting the grading; Appendix~\ref{app:r-sector-anyon-map} illustrates such an example.

The anyon permutations Eqs. \eqref{eq:aksoy-wen-third-line} and \eqref{eq:onsite_FAD_anyon}, induced by our symmetry-permuting entangler Eq. \eqref{eq:final} and the on-site unitary Eq. \eqref{eq:onsite_FAD}, respectively, generate the full $S_3$ group of FADs in Ref.~\cite{aksoy2025}.

\section{Discussion and Outlook}
\label{sec:discussion-outlook}

We constructed a symmetry-permuting finite-depth entangler connecting distinct $\Rep(D_8)$ SPT phases that cannot be connected by a symmetric entangler. This provides a microscopic realization of the distinction between fixed-charge and fixed-algebra dualities anticipated by the symTFT classification. Moreover, the circuit action on string operators reproduces the anyon permutation of the corresponding FAD, providing direct microscopic support for the symTFT description of these dualities.

Several questions remain. Although our construction is less ad hoc than the earlier $\Rep(A_4)$ example and is guided by local crossing relations for the MPO tensors, it is not yet algorithmic. Moreover, while our method  extracts the anyon permutation associated with the braided autoequivalence for the symmetry-permuting entangler via its action on  string operators,  it does not necessarily determine the braided autoequivalence itself: in general, two distinct braided autoequivalences can have the same anyon permutation action. \footnote{For symmetric entanglers with group symmetry, the analogous distinction is between the group $2$-cocycle and its slant product. For some groups the slant product determines the full cohomology class, whereas this fails in the presence of Bogomolov multipliers \cite{Pollmann_2012, Davydov_2014,kobayashi2025soft,kobayashi2025projective}.}

After a fuller account of the relation between finite-depth circuits and FADs, a natural next step would be to move beyond FADs to the more general fixed-symmetry dualities (FSDs) of Ref.~\cite{aksoy2025}. Our results suggest that FADs are  amenable to finite-depth realizations. It would be interesting to study more general duality operators which realize FSDs from our general perspective on MPO circuits and anyon permutations. 

\vspace{1cm}
\emph{Acknowledgments}: I thank Takamasa Ando for helpful discussions. I also thank Clement Delcamp for discussions clarifying the difference between the two notions of dualities. GPT-5.6 Sol  (and GPT-6 Astra, to a limited extent) was used extensively for conceptual discussions, computations, writing, and drawing diagrams.

\appendix
\section{Anyon dictionary for \texorpdfstring{$Z(\operatorname{Rep}(D_8))$}{Z(Rep(D8))}}
\label{app:anyon-dictionary}

In this appendix we summarize our conventions for the anyons of
$$
Z(\operatorname{Rep}(D_8))
\simeq
Z(\mathrm{Vec}_{D_8})$$
and give the dictionary to the notation of Ref.~\cite{aksoy2025}.

We use
\begin{equation}
D_8
=
\langle r,s
\mid
r^4=s^2=1,\;
srs=r^{-1}
\rangle .
\label{eq:appA-D8-definition}
\end{equation}
The anyons of the quantum double are labeled by pairs
\begin{equation}
([g],\rho),
\end{equation}
where $[g]$ is a conjugacy class of $D_8$ and $\rho$ is an irreducible representation of the centralizer $Z_{D_8}(g)$.

The conjugacy classes are
\begin{equation}
[e]=\{e\},
\qquad
[r^2]=\{r^2\},
\end{equation}
\begin{equation}
[r]=\{r,r^3\},
\qquad
[s]=\{s,r^2s\},
\qquad
[rs]=\{rs,r^3s\}.
\end{equation}
Their centralizers are
\begin{equation}
Z_{D_8}(e)
=
Z_{D_8}(r^2)
=
D_8,
\end{equation}
\begin{equation}
Z_{D_8}(r)
=
\langle r\rangle
\simeq
\mathbb Z_4,
\end{equation}
\begin{equation}
Z_{D_8}(s)
=
H
=
\langle r^2,s\rangle
\simeq
\mathbb Z_2\times\mathbb Z_2,
\end{equation}
and
\begin{equation}
Z_{D_8}(rs)
=
H'
=
\langle r^2,rs\rangle
\simeq
\mathbb Z_2\times\mathbb Z_2.
\end{equation}

\subsection{Representation conventions}

The irreducible representations of $D_8$ are
\begin{equation}
\mathbf  1,
\qquad
\rho_a,
\qquad
\rho_b,
\qquad
\rho_c=\rho_a\rho_b,
\qquad
\sigma,
\end{equation}
with
\begin{equation}
\rho_a(r)=1,
\qquad
\rho_a(s)=-1,
\end{equation}
\begin{equation}
\rho_b(r)=-1,
\qquad
\rho_b(s)=1,
\end{equation}
and
\begin{equation}
\sigma(r)=-iY,
\qquad
\sigma(s)=Z.
\end{equation}

For 
$$H=\langle r^2,s\rangle \simeq \ZZ_2 \times \ZZ_2,$$ 
we denote its four one-dimensional representations by
\[
\omega_1,\omega_2,\omega_3,\omega_4,
\]
with values on the generators $(r^2,s)$
\begin{equation}
\begin{array}{c|cc}
 & r^2 & s\\
\hline
\omega_1 & +1 & +1\\
\omega_2 & -1 & +1\\
\omega_3 & +1 & -1\\
\omega_4 & -1 & -1
\end{array}.
\label{eq:appA-H-irreps}
\end{equation}
Similarly, for
$$H'=\langle r^2,rs\rangle \simeq \ZZ_2 \times \ZZ_2,$$
we write
$$
\widetilde\omega_1,\ldots,\widetilde\omega_4
$$
with the same sign convention on the generators $(r^2,rs)$.

For
$$
Z_{D_8}(r)
=
\langle r\rangle
\simeq
\mathbb Z_4,
$$
we choose the character
\begin{equation}
\chi(r)=i,
\end{equation}
so the four irreducible representations are
$$
\chi^0,\chi^1,\chi^2,\chi^3.
$$

\subsection{Dictionary to Aksoy and Wen}

With these conventions, the anyon labels of Appendix C of Ref.~\cite{aksoy2025} are
\begin{center}
\begin{tabular}{c|c}
\hline\hline
Aksoy--Wen & Anyon in our notation\\
\hline
$1$ & $([e],\mathbf 1)$\\
$a$ & $([e],\rho_b)$\\
$b$ & $([e],\rho_c)$\\
$g$ & $([e],\rho_a)$\\
$c$ & $([r^2],\rho_a)$\\
$e$ & $([r^2],\rho_c)$\\
$h$ & $([r^2],\rho_b)$\\
$d$ & $([r^2],\mathbf 1)$\\
$i$ & $([e],\sigma)$\\
$m$ & $([s],\omega_2)$\\
$l$ & $([s],\omega_1)$\\
$j$ & $([rs],\widetilde\omega_2)$\\
$n$ & $([rs],\widetilde\omega_1)$\\
$k$ & $([r],\chi^0)$\\
$u$ & $([s],\omega_4)$\\
$q$ & $([s],\omega_3)$\\
$s$ & $([rs],\widetilde\omega_4)$\\
$v$ & $([rs],\widetilde\omega_3)$\\
$r$ & $([r^2],\sigma)$\\
$t$ & $([r],\chi^2)$\\
$x$ & $([r],\chi^1)$\\
$y$ & $([r],\chi^3)$\\
\hline\hline
\end{tabular}
\end{center}

In particular, the pure charges appearing in the main text are
\begin{equation}
a_{\rm AW}=([e],\rho_b),
\qquad
b_{\rm AW}=([e],\rho_c),
\qquad
g_{\rm AW}=([e],\rho_a),
\end{equation}
while
\begin{equation}
i_{\rm AW}=([e],\sigma).
\end{equation}

\subsection{The three \texorpdfstring{$\operatorname{Rep}(D_8)$}{Rep(D8)} SPT boundaries}

In the notation of Ref.~\cite{aksoy2025}, the three SPT Lagrangian algebras are
\begin{equation}
\mathcal A_{3,1}
=
1\oplus 2i\oplus a\oplus b\oplus g,
\end{equation}
\begin{equation}
\mathcal A_{3,2}
=
1\oplus 2m\oplus a\oplus c\oplus e,
\end{equation}
and
\begin{equation}
\mathcal A_{3,4}
=
1\oplus 2j\oplus b\oplus c\oplus h.
\end{equation}

In our notation these become
\begin{equation}
\mathcal A_{3,1}
=
([e],\mathbf 1)
\oplus
2([e],\sigma)
\oplus
([e],\rho_a)
\oplus
([e],\rho_b)
\oplus
([e],\rho_c),
\label{eq:appA-A31}
\end{equation}
corresponding to the trivial SPT,
\begin{equation}
\mathcal A_{3,2}
=
([e],\mathbf 1)
\oplus
2([s],\omega_2)
\oplus
([e],\rho_b)
\oplus
([r^2],\rho_a)
\oplus
([r^2],\rho_c),
\label{eq:appA-A32}
\end{equation}
corresponding to the
\[
H=\langle r^2,s\rangle
\]
SPT, and
\begin{equation}
\mathcal A_{3,4}
=
([e],\mathbf 1)
\oplus
2([rs],\widetilde\omega_2)
\oplus
([e],\rho_c)
\oplus
([r^2],\rho_a)
\oplus
([r^2],\rho_b),
\label{eq:appA-A34}
\end{equation}
corresponding to the
\[
H'=\langle r^2,rs\rangle
\]
SPT.

The non-onsite FAD realized by the MPU constructed in this work is the third involution in Eq.~(C7) of Ref.~\cite{aksoy2025},
\begin{equation}
(i,m)(n,k)(u,r)(v,t)(b,c)(g,e).
\label{eq:appA-target-FAD-AW}
\end{equation}
In the notation above,
\begin{equation}
\boxed{
\begin{aligned}
([e],\sigma)
&\leftrightarrow
([s],\omega_2),\\
([rs],\widetilde\omega_1)
&\leftrightarrow
([r],\chi^0),\\
([s],\omega_4)
&\leftrightarrow
([r^2],\sigma),\\
([rs],\widetilde\omega_3)
&\leftrightarrow
([r],\chi^2),\\
([e],\rho_c)
&\leftrightarrow
([r^2],\rho_a),\\
([e],\rho_a)
&\leftrightarrow
([r^2],\rho_c).
\end{aligned}}
\label{eq:appA-target-FAD-ours}
\end{equation}
It exchanges
\begin{equation}
\mathcal A_{3,1}
\longleftrightarrow
\mathcal A_{3,2}.
\end{equation}

The second generator used in the main text is induced by the group automorphism
\begin{equation}
r\mapsto r^{-1},
\qquad
s\mapsto rs,
\label{eq:appA-onsite-automorphism}
\end{equation}
which exchanges
\begin{equation}
\mathcal A_{3,2}
\longleftrightarrow
\mathcal A_{3,4}
\end{equation}
while fixing the trivial SPT boundary. Together their anyon permutations generate the $S_3$ group of FADs discussed in Sec.~\ref{subsec:repD8-anyon-action}.

\section{Crossing tensors from action tensors}
\label{app:action_tensors}

Suppose an FDLU with MPO tensors $M$ (with bond dimension $D_M$) maps an SPT state with MPS tensors $A$ (with bond dimension $D_A$) and  action tensors $\Lambda$ to another SPT state with tensors $A'$ (with bond dimension $D_{A'}$) and $\Lambda'$, i.e. 
\begin{equation}
    A'^{m} =  \sum_n M^{m,n} \otimes A^n, 
\end{equation}
as is the case at hand. Then, using the definition of action tensors \cite{Garre_Rubio_2023}
\begin{align}
    T_{S_a}^{m,n}  \otimes A^n = (\Lambda_a^i)^\dagger  A^m \Lambda_a^i 
\end{align}
we obtain 
\begin{align}
    \Lambda'^i_a
= \left(I_{D_M}\otimes \Lambda_{a}^i \right)
\left(\mathcal{X}_a \otimes I_{D_A}\right).
\label{eq:action_crossing}
\end{align}

We now apply this to our example. For the trivial MPS, the action tensors for the one-dimensional symmetry lines are
\begin{equation}
\Lambda_{\mathbf 1}^{(0)}
=
\Lambda_{\rho_a}^{(0)}
=
\Lambda_{\rho_b}^{(0)}
=
\Lambda_{\rho_c}^{(0)}
=
(1).
\end{equation}
For the two-dimensional symmetry line $\sigma$, the two fusion channels may be chosen as
\begin{equation}
\Lambda_\sigma^{(0),0}
=
\begin{pmatrix}
1&0
\end{pmatrix},
\qquad
\Lambda_\sigma^{(0),1}
=
\begin{pmatrix}
0&1
\end{pmatrix}.
\label{eq:appB-product-sigma-actions}
\end{equation}

For the $H$-SPT MPS, the invertible symmetry action tensors are
\begin{equation}
\Lambda_{\mathbf 1}^{(\omega)}=I,
\qquad
\Lambda_{\rho_b}^{(\omega)}=I,
\end{equation}
and
\begin{equation}
\Lambda_{\rho_a}^{(\omega)}=X,
\qquad
\Lambda_{\rho_c}^{(\omega)}=X.
\label{eq:appB-SPT-invertible-actions}
\end{equation}
For the noninvertible line,
\begin{equation}
\Lambda_\sigma^{(\omega),0}
=
\langle0|\otimes Z
=
\begin{pmatrix}
1&0&0&0\\
0&-1&0&0
\end{pmatrix},
\end{equation}
and
\begin{equation}
\Lambda_\sigma^{(\omega),1}
=
\langle1|\otimes Y
=
\begin{pmatrix}
0&0&0&-i\\
0&0&i&0
\end{pmatrix}.
\label{eq:appB-SPT-sigma-actions}
\end{equation}

Eq. \eqref{eq:action_crossing} then gives: 
\begin{equation}
\mathcal X_{\mathbf 1}=I,\qquad
\mathcal X_{\rho_b}=I,\qquad
\mathcal X_{\rho_a}=X,\qquad
\mathcal X_{\rho_c}=X.
\label{eq:appB-invertible-zippers}
\end{equation}
and 
\begin{equation}
\mathcal X_\sigma
=
|0\rangle\langle0|\otimes Z
+
|1\rangle\langle1|\otimes Y.
\label{eq:appB-sigma-zipper}
\end{equation}
These are the crossing tensors used in the main text.

\section{Microscopic transformation of the \texorpdfstring{$([r],\chi^0)$}{([r],chi0)} anyon sector}
\label{app:r-sector-anyon-map}

In this Appendix we illustrate explicitly how the local MPU tensor determines the transformation of a nontrivially graded anyon. We consider
$$
([r],\chi^0)\in \rm{Irr} (Z(\Rep(D_8))),
$$
and show directly from the endpoint pulling-through equations that
$$
([r],\chi^0)
\longmapsto
([rs],\widetilde\omega_1).
$$

Here

$$
[r]=\{r,r^3\},
\qquad
[rs]=\{rs,r^3s\},
$$

and

$$
Z_{D_8}(rs)
=
H'
=
\langle r^2,rs\rangle
\simeq \mathbb Z_2\times\mathbb Z_2.
$$

We denote by \(\widetilde\omega_1\) the trivial one-dimensional representation of \(H'\),

$$
\widetilde\omega_1(r^2)=
\widetilde\omega_1(rs)=1.
$$

\subsection{Graded and symmetry-adapted endpoint bases}

As a $D_8$-graded representation, the anyon $([r],\chi^0)$ is
$$
W=W_r\oplus W_{r^3},
\qquad
\dim W_r=\dim W_{r^3}=1.
$$
Upon forgetting the grading, the induced \(D_8\)-representation decomposes as
$$
W\simeq \mathbf 1\oplus\rho_a.
$$
Thus there are two useful bases for the endpoint tensor: a homogeneous basis adapted to the grading $r,r^3$, and an inhomogeneous basis adapted to the symmetry irreps $\mathbf 1,\rho_a$.

Since the right-translation operator $R_g$ has grade $g$, a convenient representative of a homogeneous endpoint component may be written  $R_gD,$  where $D$ is diagonal and hence grade neutral.

We choose
\begin{equation}
\Gamma_r=R_rD_r,
\qquad
\Gamma_{r^3}=R_{r^3}D_{r^3},
\end{equation}
with physical basis ordered as
$$
(e,r^2,s,r^2s\mid r,r^3,rs,r^3s)
$$
and
$$
D_r=
\rm{diag}(1,i,1,i\mid0,0,0,0),
$$
$$
D_{r^3}
=
\rm{diag}(i,1,i,1\mid0,0,0,0).
$$
Equivalently, writing
$$
u=r^2,\qquad v=s,\qquad
h=u^mv^n\in H=\langle u,v\rangle,
$$
their nonzero diagonal entries are
$$
d_r(h)=i^m,
\qquad
d_{r^3}(h)=i(-i)^m.
$$
These diagonal factors are a convenient microscopic gauge choice and do not carry  topological information. In particular, one could instead begin from the bare representatives $R_r,R_{r^3}$; the transformed operators would then contain more complicated grade-neutral diagonal factors, while the output grades derived below would be unchanged.

Passing to the symmetry-adapted basis gives, up to normalization,
\begin{align}
\Gamma_{\mathbf 1}
=
\Gamma_r+\Gamma_{r^3},
\qquad
\Gamma_{\rho_a}
=
\Gamma_r-\Gamma_{r^3}.
\end{align}
The two components are no longer homogeneous with respect to the $D_8$-grading, but they now terminate definite symmetry lines and can therefore be pulled through using the corresponding crossing tensors.

For the entangler considered in the main text,
$$
\mathbf 1\longmapsto\mathbf 1,
\qquad
\mathcal X_{\mathbf 1}=I,
$$
whereas
$$
\rho_a\longmapsto\rho_c,
\qquad
\mathcal X_{\rho_a}=X.
$$

\subsection{Endpoint pulling-through equations}

Let

$$
M=
\sum_{g,h\in D_8}
|g\rangle\langle h|\otimes M^{g,h}
$$

be the local MPU tensor. The endpoint pulling-through equations for the two symmetry components take the form
\begin{align}
\sum_q
(\Gamma_{\mathbf1})^g{}_q
M^{q,h}
=
\sum_t
M^{g,t}
(\widetilde\Gamma_{\mathbf1})^t{}_h, \nonumber \\ 
\sum_q
(\Gamma_{\rho_a})^g{}_q
M^{q,h}X
=
\sum_t
M^{g,t}
(\widetilde\Gamma_{\rho_c})^t{}_h.
\label{eq:app-r}
\end{align}
From the relation Eq. \eqref{eq:X_relation} we see that the crossing matrix $X$ acts, at the level of physical grading, as an $r^2$-shift of the second physical index.

The appearance of the additional factor of $s$ in the final conjugacy-class transformation is more subtle. It follows from the relation between the odd and even blocks of $M$. Substitution of the above endpoint gauge into Eq.~\eqref{eq:app-r} gives
\begin{align}
d_r(gr)M^{gr,h}
+
d_{r^3}(gr^3)M^{gr^3,h} \nonumber \\ 
=
\alpha_h M^{g,hrs}
+
\beta_h M^{g,hr^3s},
\label{eq:app-r-sum}
\end{align}
while Eq.~\eqref{eq:app-r}, together with Eq.~\eqref{eq:X_relation}, gives
\begin{align}
\left[
d_r(gr)M^{gr,h}
-
d_{r^3}(gr^3)M^{gr^3,h}
\right]X \nonumber \\
=
i\alpha_h M^{g,hrs}
-
i\beta_h M^{g,hr^3s}.
\label{eq:app-r-difference}
\end{align}
Here, 
\begin{equation}
\alpha_h
=
\frac{-1+i(-1)^m}{\sqrt2},
\qquad
\beta_h
=
\frac{1+i(-1)^m}{\sqrt2},
\end{equation}
for \(h=u^mv^n\in H\).
We define the corresponding diagonal operators
\begin{align}
    D_\alpha
=
\frac1{\sqrt2}
\rm{diag}(
-1+i,-1-i,-1+i,-1-i
\mid
0,0,0,0),
\end{align}
and
\begin{align}
D_\beta
=
\frac1{\sqrt2}
\rm{diag}(
1+i,1-i,1+i,1-i
\mid
0,0,0,0).
\end{align}

Equations~\eqref{eq:app-r-sum} and \eqref{eq:app-r-difference} therefore identify the transformed symmetry components as
\begin{align}
\widetilde\Gamma_{\mathbf1}
=
R_{rs}D_\alpha
+
R_{r^3s}D_\beta,
 \nonumber \\ 
\widetilde\Gamma_{\rho_c}
=
iR_{rs}D_\alpha
-
iR_{r^3s}D_\beta.
\end{align}
Returning from the \(\mathbf1,\rho_c\) basis to a homogeneous grading basis gives
\begin{align}
    \frac12
\left(
\widetilde\Gamma_{\mathbf1}
-i\widetilde\Gamma_{\rho_c}
\right)
=
R_{rs}D_\alpha, \nonumber \\ 
\frac12
\left(
\widetilde\Gamma_{\mathbf1}
+i\widetilde\Gamma_{\rho_c}
\right)
=
R_{r^3s}D_\beta.
\end{align}

The two transformed homogeneous components therefore have grades
$$rs,\qquad r^3s.
$$
Hence the conjugacy-class label transforms as
$$
[r]=\{r,r^3\}
\longmapsto
[rs]=\{rs,r^3s\}.
$$
\subsection{Centralizer representation and the anyon}
Having determined the transformed grading, it remains to identify the representation of the stabilizer of \(rs\). Since
$$
Z_{D_8}(rs)
=
H'
=
\langle r^2,rs\rangle,
$$
Our transformed ordinary \(D_8\)-representation is
$$
\mathbf1\oplus\rho_c.
$$
Both $\mathbf 1$ and $\rho_c$ restrict to the trivial representation on $H'$ since $\rho_c(r^2) = \rho_c(rs) = 1$. 
The centralizer representation is therefore precisely $\widetilde\omega_1$ in Appendix \ref{app:anyon-dictionary}, and we conclude
\begin{equation}
([r],\chi^0)
\longmapsto
([rs],\widetilde\omega_1)
\end{equation}
as the anyon transformation induced by the entangler.

\bibliography{ref}

\end{document}